\documentclass[10pt]{llncs}
\usepackage[hidelinks,unicode,linktoc=all,pdfwindowui,pdfpagemode=FullScreen]{hyperref}
\hypersetup{
  colorlinks   = true, 
  urlcolor     = blue, 
  linkcolor    = blue, 
  citecolor   = blue 
}

\usepackage[T1]{fontenc}
\usepackage{etoolbox}
\usepackage{graphicx}

\usepackage{amssymb}
\usepackage{pifont}
\newcommand{\cmark}{\ding{51}}%
\newcommand{\xmark}{\ding{55}}%

\title{Securing Stack Smashing Protection in WebAssembly Applications}
\author{
  Quentin Michaud \inst{1,2}
  \and Yohan Pipereau \inst{2}
  \and Olivier Levillain \inst{2}
  \and Dhouha Ayed \inst{1}
}

\institute{
  Thales Group, Palaiseau, France\\
  \email{firstname.lastname@thalesgroup.com}
  \and SAMOVAR, Télécom SudParis, Institut Polytechnique de Paris, Palaiseau, France\\
  \email{firstname.lastname@telecom-sudparis.eu}
}

\begin{document}

\maketitle

\begin{abstract}
WebAssembly is an instruction set architecture and binary format standard, designed for secure execution by an interpreter.
Previous work has shown that WebAssembly is vulnerable to buffer overflow due to the lack of effective protection mechanisms.

In this paper, we evaluate the implementation of Stack Smashing Protection (SSP) in WebAssembly standalone runtimes, and uncover two weaknesses in their current implementation.
The first one is the possibility to overwrite the SSP reference value because of the contiguous memory zones inside a WebAssembly process.
The second comes from the reliance of WebAssembly on the runtime to provide randomness in order to initialize the SSP reference value, which impacts the robustness of the solution.

We address these two flaws by hardening the SSP implementation in terms of storage and random generator failure, in a way that is generalizable to all of WebAssembly.
We evaluate our new, more robust, solution to prove that the implemented improvements do not reduce the efficiency of SSP.
\end{abstract}

\section{Introduction}
\label{introduction}

WebAssembly~\cite{rossberg_webassembly_2019,haas_bringing_2017} has been created as a fast and secure-by-design answer to the always increasing need for complex computation in browsers, such as 3D rendering~\cite{webgl}, 3D model parsing~\cite{wasm_model_parsing}, gaming~\cite{wasm_games}, hardware emulation~\cite{tinyemu} or physics workloads (e.g.~computational fluid dynamics~\cite{sakuta_computational_2024}).

The success of WebAssembly as a portable Instruction Set Architecture (ISA) and binary format has prompted its adoption in many applications besides browsers.

Today, we can find WebAssembly in smart contracts~\cite{smart_contracts,mccallum_diving_2019}, embedded devices~\cite{gurdeep_singh_warduino_2019}, secure plugins~\cite{narayan_gobi_2019,noauthor_extism_nodate}, Function as a Service (FaaS) platforms~\cite{faas,kjorveziroski_webassembly_2023}, or as a standalone runtime~\cite{clark_standardizing_nodate}.
The latter has a huge impact on the cloud world and the computing world in general.
Some see in the flexibility of WebAssembly a universal binary format that could be distributed seamlessly across operating systems and hardware architectures.
It also appears in various cloud-related projects and is considered as an alternative to Linux-based containers~\cite{noauthor_wasmcncf_2024,noauthor_wasmcloud_nodate}, promising to be more portable, lightweight and secure.

WebAssembly claims strong security.
By default, it provides sandboxing between different WebAssembly instances and between WebAssembly and its host. 
It also enforces control-flow integrity, and protection against code reuse attacks. 
However, the security of WebAssembly has been challenged in several works~\cite{qrs::21::stievenart,usenix_sec::20::lehmann}.
First, WebAssembly offers weak protection against memory corruption attacks compared to native binaries.
Some vulnerabilities, such as stack-based buffer overflows, have been present in native binaries for a long time, but are mitigated with mechanisms such as Stack Smashing Protection (SSP).
This protection was initially absent in WebAssembly~\cite{webassembly_security_doc}.
Second, differences in design between WebAssembly and native binaries make the former vulnerable to attacks that are not possible in native binaries.
One example is the corruption of heap data using a stack-based buffer overflow.

Stack Smashing Protection has been implemented in WebAssembly after the publication of the papers discussed in the previous paragraph.
In this paper, we propose the following contributions: (i) a thorough analysis of SSP in WebAssembly;
(ii) some proofs of concept to confirm the weaknesses of the current implementation;
(iii) the implementation of a more robust SSP mechanism in LLVM~\cite{llvm} and \texttt{wasi-libc};\footnote{https://github.com/WebAssembly/wasi-libc/} (iv) an evaluation of our solution.

We open-source all our code contributions: the implementation of SSP in the WebAssembly target of the LLVM compiler;\footnote{https://github.com/ThalesGroup/llvm-project/tree/new-wasm-ssp} modifications to \texttt{wasi-libc};\footnote{https://github.com/ThalesGroup/wasi-libc/tree/new-wasm-ssp} adaptation of \emph{CookieCrumbler}~\cite{ifip::18::cookiecrumbler} (a tool used to assess the robustness of SSP implementations) to WebAssembly; and our proofs of concept.\footnote{https://github.com/mh4ck-Thales/Robust-SSP-in-Wasm}

The following if this paper is structured as follows.
First, Section~\ref{background} and~\ref{motivation} presents necessary background and motivation for this work. 
Then, Section~\ref{security-analysis-of-webassembly-ssp} contains our security analysis of WebAssembly SSP and our remediation proposals.
Finally, Section~\ref{evaluation} provides an evaluation of our work and Section~\ref{conclusion} concludes and gives some perspective for future work.

\section{Background}
\label{background}

We start by giving a brief introduction to buffer overflows and Stack Smashing Protection.
We also provide a quick background on WebAssembly and its inner workings.

\subsection{Buffer overflow and Stack Smashing Protection}

\subsubsection{Buffer Overflow.}

Buffer overflows are an old and well-known vulnerability~\cite{one06phrack}.
They occur when a program stores more data in a buffer than the buffer may hold.
Writing to memory out of buffer bounds leads to the corruption of memory adjacent to the buffer.

Buffer overflows may also happen during the execution of a WebAssembly program.

\subsubsection{Stack Smashing Protection.}

\emph{Stack Smashing Protection} (SSP), also known as \emph{stack canaries} or \emph{stack cookies}~\cite{usenix_sec::98::stackguard} is a defense mechanism available to prevent exploitation of stack-based buffer overflows.
SSP provides a detection mechanism for stack-based buffer overflows and terminates the execution of the program after the current function is executed.

At program start time, the program initializes a random reference value (named \emph{canary} or \emph{cookie}) and writes it in a memory zone, preferably where overwrite is made impossible, or at least difficult.

Each time a function is called, the function prologue is executed which creates a new stack frame and copies the canary reference value in the stack, in a dedicated variable, the \emph{stack canary} located after the buffer.
The function epilogue checks this value against the canary reference value stored in safe memory.
If the stack canary is different from the reference value, it means that the stack canary has been overwritten and that a stack-based buffer overflow has occurred.
In this case, a specific function is called to terminate the process.

Stack Smashing Protection is implemented in two different code bases.
The initialization of the reference value and the function called when the stack canary is overwritten is provided in the language standard library (e.g.~the GNU C standard library or the musl C standard library).
The generation of the specific function prologue and epilogue for setting up and verifying the integrity of canaries is implemented in the compiler.

\subsection{WebAssembly}

\subsubsection{Overview.}

WebAssembly (commonly abbreviated as Wasm) is a binary format, designed to be compact, easy to parse and fast at execution.
A WebAssembly file, containing a WebAssembly program, is named a \emph{module}.
An \emph{instance} is a module being executed in a runtime.
WebAssembly is also an Instruction Set Architecture (ISA), designed as a stack-based virtual machine.
It was designed to be fast and secure by design.

A lot of programming languages can be compiled to WebAssembly, with several compilers for various languages and environments.
Among the commonly used languages are C, C++, Rust, Go, and AssemblyScript, and support is constantly growing.
Alongside the source code, compilers need to add their own code and libraries for adapting the program to its host environment. 
For example, browser-based WebAssembly includes specific JavaScript bindings to allow WebAssembly to interact with the functionalities of the browser.
This means that such a WebAssembly binary will not be able to run in a different environment, e.g.~in standalone mode on a server.

\subsubsection{Execution model.}
\label{execution-model}

WebAssembly bytecode is executed using a stack-based Virtual Machine (VM).
This means that each instruction gets input operands by popping values off a stack, and pushes its eventual results on this stack referred as the \emph{evaluation stack}.
There are no registers in the WebAssembly virtual machine.

The WebAssembly bytecode is located in a specific memory managed by the virtual machine, that is read to execute instruction, but that is not directly accessible by the program.

In this architecture, the call stack is also stored in a dedicated memory. This means that the return address for each function call is saved separately from the linear memory, thus implicitly implementing backwards-edge control-flow integrity (i.e.~integrity when resuming the caller execution) and offering a strong protection against control-flow hijacking.

Moreover, WebAssembly also implicitly enforce forward-edge control-flow integrity (i.e.~integrity when calling a new function) by using function tables.
Function tables list which functions are present in the binary, where they are located in the code and what arguments they expect.
Only functions present in the table can be called, ensuring that arbitrary assembly code cannot be executed.
However, to implement function pointers, WebAssembly has an \texttt{indirect\_call} instruction, which is using data from the linear memory to determine which function to call in the function table.
The signature of the function in the WebAssembly code must match the signature of the function in the function table, but it is still possible for an attacker to corrupt the data in the linear memory to divert the control flow by calling another function with the same signature.

\subsubsection{Memory model.}

\begin{figure}
	\centering
	\includegraphics[width=.9\textwidth]{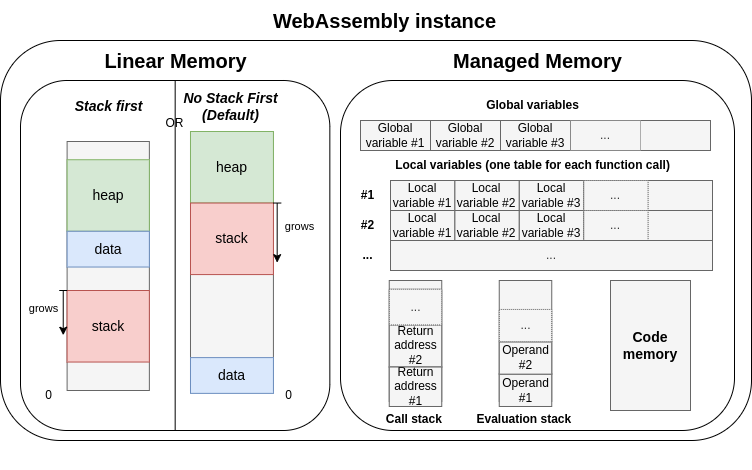}
	\caption{The memory layout of a WebAssembly virtual machine}
	\label{fig::wasm-mem}
\end{figure}

The WebAssembly virtual machine relies on multiple memory regions which are represented in Figure~\ref{fig::wasm-mem}, and summarized below.

The \emph{managed code memory} contains the WebAssembly program code.
It is only accessible by the VM, so the WebAssembly code cannot read or modify it.

The \emph{managed call stack} contains return addresses.
These return addresses are of WebAssembly's \texttt{i32} type, which is used as the type for memory pointers and addresses.\footnote{A proposed extension of WebAssembly is using the \texttt{i64} type to address the memory, but this extension is not addressed in this paper.}
It is used to keep track of all the ongoing function calls, while preventing control-flow hijacking based on return address overwrite.

The \emph{managed evaluation stack} is used to give parameters to instructions and to store their results.
This stack can hold the four WebAssembly basic types, i.e.~\texttt{i32}, \texttt{i64}, \texttt{f32} and \texttt{f64} that are respectively integers and floating point numbers encoded on 32 or 64 bits.

The \emph{linear memory} is used to store non-scalar types, e.g.~strings, arrays, or lists.
This linear memory is a single contiguous memory segment with no notion of memory permissions.
As such, all memory in the linear memory is readable and writable.
Linear memory does not use any address randomization mechanisms, such as Address Space Layout Randomization (ASLR) or Position-Independent Executables (PIE), which are supported by all major operating systems.
The management of this memory is left to the program, but for most programming languages and their respective compilers, the structure used is the same as the one widely used in native binaries, which includes a stack, a heap and a data zone for static or predetermined values.
These memory zones contain most of the data used by the program, the data being distributed between the different zones according to the source code and the compiler used.
The situations relevant to this paper will be further described in Section~\ref{memory-layout}.

The \emph{WebAssembly local and global variables} are another memory mechanism.
As for the evaluation stack, they are restricted to the four WebAssembly basic types.
The scope of global variables is the entire module, while local variables are only accessible by the function being executed.
These variables are manipulated through dedicated instructions and are stored in a specific table that is not accessible from the linear memory.
It is however important to notice that current toolchains do \emph{not} usually map local and global variables found in programming languages onto WebAssembly local and global variables.

\subsection{Standalone WebAssembly}

\subsubsection{WASI.}

By design, WebAssembly does not provide access to the host environment in which the WebAssembly program is executed.
It can only be performed using functions provided by the WebAssembly runtime, that will then interact with the host environment as requested and store the results in linear memory or in the evaluation stack, as an internal WebAssembly function would do.
It is up to the runtime to implement or not these special functions.
In order for a WebAssembly binary to work with a large panel of runtimes and host environments, standardizing such special functions was needed.

For standalone WebAssembly, this led to the creation of the WebAssembly System Interface (WASI).
It is composed of a set of modular standards regrouped around different use cases: filesystem, random, sockets, etc.

One of the main inspirations for the design of WASI is the POSIX standards.
This brings the development of WebAssembly applications using WASI very close to Linux ones.
Indeed, a big part of Linux applications can be recompiled without any changes using a WASI compilation toolchain.
This proximity allows us to easily compare the implementations of security mechanisms between native and WebAssembly binaries.
It gives security researchers a baseline to compare against when designing or evaluating new security mechanisms for WebAssembly.

WASI~\cite{wasi_dev} is still evolving as a standard, but it is already widely used.
Two main versions of WASI exist to this day: WASI preview 1 (WASIp1), released in late 2020 and WASI preview 2 (WASIp2), released in the beginning of 2024.
There is ongoing effort to implement WASIp2 in most toolchains that are supporting WASIp1, but at the time of writing, most WebAssembly binaries still use WASIp1.
This paper along with its proposed proofs of concept is using WASIp1, as most research was done before the publication of WASIp2.

\subsubsection{Memory layout.}
\label{memory-layout}

WebAssembly compilers are leveraging the linear memory to create a memory layout with three memory zones: a stack, a heap and a zone for static data.
These three zones can be arranged in several ways in memory, and in practice different WebAssembly compilers made different choices resulting in different layouts.
For the purpose of this article, we focus on the two layouts available with the LLVM toolchain, named \emph{stack-first} and \emph{no-stack-first}.
These layouts are represented in Figure~\ref{fig::wasm-mem}.

LLVM default memory layout, \emph{no-stack-first}, puts the fixed-size data zone at the lowest addresses, followed by the stack growing downwards, and the heap growing upwards.
Due to the lack of memory separation in WebAssembly, a stack overflow (i.e.~a situation where the stack grows too much and collides with another memory region) in this layout silently corrupts data in the data zone.
Because of this drawback, Rust developers introduced the \emph{stack-first} memory layout which puts the stack growing downwards in lower addresses, with the data zone and the heap at higher address.
This layout makes the WebAssembly runtime crash in case of stack overflow, because the stack will grow until it reaches non-existent (negative) memory addresses.
This crash happens without overwriting other data first, hence indicating that a stack overflow occurred, and removing the possibility of undefined behavior because of stack overflow.
As of today, It has been adopted by default in Rust, in Zig, and LLVM is discussing to make it a default.

\section{Motivation and related work}
\label{motivation}

\subsection{WebAssembly lack of memory protection}

WebAssembly security has already been studied in several works.
Lehmann et al.~\cite{usenix_sec::20::lehmann} conduct an in-depth security analysis of the WebAssembly linear memory, and how it is used by programs compiled from various languages.
It shows that common memory protections are missing from WebAssembly, and demonstrates how this lack makes code less secure than when compiled to a native binary.
It concludes by discussing some mitigations, including the proposition to port protections provided by compilers to WebAssembly.
One of these mitigations is Stack Smashing Protection.
Our first proof of concept, corresponding to the \texttt{-no-ssp} files in our artifact repository, is inspired by their work and proves that buffer overflows in standalone WebAssembly are exploitable in practice.

However, the effectiveness of Stack Smashing Protection in WebAssembly is not guaranteed due to the great differences between WebAssembly and native binaries, and the security of its implementations have not been assessed yet.
Other propositions of mitigation mainly require significant work in the WebAssembly specifications and its extensions, and thus have not been adopted yet.

In~\cite{qrs::21::stievenart}, Stiévenart et al.~study a corpus of thousands of C programs vulnerable to stack-based~\cite{cwe-121} and heap-based buffer overflows~\cite{cwe-122}.
They compare the behavior of these programs when they are compiled as x86 binaries with state-of-the-art protections (including Stack Smashing Protection) and WebAssembly binaries, that did not have Stack Smashing Protection at the time of the study.
They observe that x86 binaries are subject to many crashes, for the most part triggered by SSP. On the contrary, WebAssembly binaries are continuing execution after the buffer overflow and memory corruption most of the time.

The difference is attributed to the absence of SSP in WebAssembly binaries, which allows an attacker to exploit buffer overflows in a stealthier fashion.
This means that WebAssembly binaries are more vulnerable to memory corruption due to buffer overflows than native ones.
At least, it means that WebAssembly binaries can see their internal memory corrupted and their data integrity violated.
In the worst case, it may be the enabler of more complex and dangerous attacks on WebAssembly (such as attacking the WebAssembly VM), as exemplified by~\cite{usenix_sec::20::lehmann}.

Since Stiévenart et al. work, SSP has been implemented in a subset of WebAssembly using LLVM and \texttt{wasi-libc}.
This means that there is no SSP available in non-WASI WebAssembly binaries, such as in the browser or depending on Node.js.
However, it could still be implemented in toolchains of these other environments using our work as the base for a secure implementation.

Zhang et al.~\cite{zhang_vmcanary_2023} propose \emph{VMCanary}, an alternative implementation of SSP for all of WebAssembly.
However, \emph{VMCanary} relies on an extension of the ISA and thus is non-standard, making it incompatible with current WebAssembly runtimes and tooling.
On the contrary, our work is based on the existing implementation in LLVM and \texttt{wasi-libc}, building on a solution which is fully compliant with the WebAssembly specifications.
Our solution has no adherence with any WebAssembly tooling, and its principles can be extended to other toolchains without breaking compatibility.

These papers conclude that WebAssembly is lacking protections that are present in native binaries.
Some security features are included in the design of WebAssembly, but there are no guarantees that they fulfill the role of the protections that are missing.
The introduction of Stack Smashing Protection on the WebAssembly world can be seen as an improvement, but its effectiveness has not been assessed yet.

\subsection{Global impact of memory corruption and protections}

In addition to assessing the possibilities of memory corruption in WebAssembly, Lehmann et al.~\cite{usenix_sec::20::lehmann} analyze the impacts of such corruption.
Their work places buffer overflow vulnerabilities as a way to potentially gain further, more impactful attack primitives.
For example, considering a program that reads and writes from and to different files, overwriting the memory contents may allow the attacker to modify a filename and thus trigger an arbitrary file read or write.

Another possibility of exploit is using restricted control flow hijacking, by abusing the \texttt{call\_indirect} instruction of WebAssembly.
This instruction allows WebAssembly to support function pointers, which are required when the compiler cannot statically determine the exact function to call (e.g.~callback functions, dynamic methods in object-oriented programming).
This makes the implicitly enforced control-flow integrity in WebAssembly weaker in the case of indirect calls.
As a result, the attacker may be able to control the function that will be called, and thus control the code that will be executed.

Hilbig et al.~\cite{hilbig_empirical_2021} study a dataset of more than 8,000 WebAssembly binaries collected from various sources in late 2020.
Among other research questions, they investigate which tools and source languages are used to produce WebAssembly binaries.
This question is more and more relevant as the popularity of WebAssembly grows and WebAssembly binaries are increasingly used in new domains.
More specifically, as the tools and use cases for WebAssembly diversify, the work needed to spread the new security propositions for WebAssembly becomes longer and longer.

One of the findings of Hilbig et al.~is that 64.2\% of WebAssembly binaries are written in C or C++, which are memory-unsafe languages.
This strongly suggests that the work on assessing memory safety in WebAssembly is relevant.
Furthermore, it underlines the importance of Stack Smashing Protection for the global security of WebAssembly binaries and the WebAssembly ecosystem.

Another finding is that nearly 80\% of all collected binaries are compiled with the help of the LLVM toolchain.
Thus, implementing a security mechanism in LLVM, such as Stack Smashing Protection, would allow introducing increased protection in most WebAssembly programs without additional engineering efforts.

\subsection{Potential weaknesses in SSP implementations}
\label{sec:potention-weaknesses-ssp}

Implementing SSP does not mean that a binary is fully protected against stack-based buffer overflows.
SSP can be bypassed when the underlying assumptions are not met.
Indeed, weak implementations of SSP allow an attacker to target the SSP mechanism in order to exploit a stack-based buffer overflow undetected.

Bierbaumer et al.~\cite{ifip::18::cookiecrumbler} conduct an analysis of the implementation of SSP across various platforms (OS, architectures and libraries) to identify potential implementation weaknesses.
They propose a list of security properties that robust SSP implementations should satisfy, and a framework named \emph{CookieCrumbler} to automatically assess the implementations.

The authors assume a buffer overflow that is contiguous and located from a buffer in the stack.
This means that the overflow does not allow the attacker to skip some bytes in memory.
The security properties that robust SSP implementations should satisfy are as follows:

\begin{enumerate}
\item[\textbf{P1}] The canary value placed behind user-controlled buffers must be unknown to the attacker.
\item[\textbf{P2}] The reference value is placed at a location in memory that is distinct from the location of canaries and ideally mapped read-only.
\item[\textbf{P3}] If a canary is corrupted, the program execution terminates immediately (or as soon as possible) without accessing any attacker controlled data.
\end{enumerate}

The main goal of Bierbaumer et al.~was to prove these properties wrong due to implementation weaknesses.
Their findings show that the robustness or SSP implementations is heterogeneous, and that some implementations are indeed vulnerable, allowing an attacker to completely bypass the protection.
Making the same analysis for the implementation of SSP in WebAssembly is interesting, as no such work has been done to the best of our knowledge.

In addition, the work of Bierbaumer et al.~was mainly targeting x86 binaries, alongside a few other results on other platforms such as ARM or PowerPC.
The inner workings of these native platforms are very far from the one of WebAssembly.
Therefore, the implementation of Stack Smashing Protection may differ a lot from the ones of native platforms, and the evaluation of the security of such an implementation is even more relevant.

\section{Security analysis of WebAssembly SSP}
\label{security-analysis-of-webassembly-ssp}

\subsection{Description of existing WebAssembly SSP and methodology}

The implementation of SSP cannot be uniform across the whole ecosystem of WebAssembly.
More precisely, an SSP implementation in WebAssembly relies on three elements, that are dependent on the target use:

\begin{itemize}
  \item The \emph{compiler}, that will provide the code for loading and checking the canaries.
  \item A \emph{library}, that will provide the code for initializing the canary reference value and the abort function that is called if a canary is overwritten.
  \item The \emph{host environment}: by nature, SSP needs randomness, that WebAssembly cannot provide by itself, so it is reliant on the host and on the way it can access or request resources from the host.
\end{itemize}

To the best of our knowledge, there is only one existing implementation of SSP in WebAssembly.
This implementation relies on both LLVM (providing the compiler) and \texttt{wasi-libc} (a C standard library targeting WASI).
As such, it is restricted to standalone WebAssembly.

In order to assess the robustness of the Stack Smashing Protection implementation, we use the properties introduced in Section~\ref{sec:potention-weaknesses-ssp}.
These criteria can be evaluated independently.
We use several methods to assess each of the properties, including source code analysis, disassembly of compiled binaries, and the \emph{CookieCrumbler} tool provided by the authors.

\subsection{Evaluating the generation of canaries}

We first assess whether the reference value is unknown to the attacker (property \textbf{P1}).
Reference values are generated using (pseudo-)randomness.
However, not all randomness guarantees a complete unpredictability.
Furthermore, one may wonder if the attacker can alter the generation of randomness, and thus compromise the generation of the reference value.

In standalone WebAssembly, the randomness is provided using WASI.
At the time of writing, the \texttt{wasi-libc} only supports WASIp1.
In this version, randomness can be acquired from the host using the \texttt{random\_get} function.
\texttt{random\_get} is able to return an error code if it is not able to provide randomness.
In the following paragraphs, we detail how this method changes across the different underlying host platforms.

\begin{figure}
\centering
\includegraphics[width=.9\textwidth]{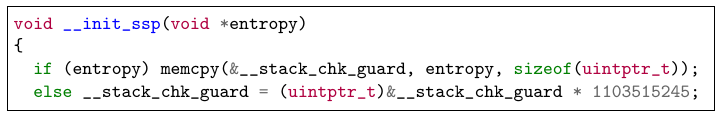}
\caption{Extract of the \texttt{init\_ssp} C function}
\label{fig:source-code-init-ssp}
\end{figure}

We can first assess the behavior of \texttt{wasi-libc} if \texttt{random\_get} returns an error code.
The code initializing the reference value is present in the \texttt{init\_ssp} function, whose relevant extracts of the source code is available in Figure~\ref{fig:source-code-init-ssp}.

In this listing, the \texttt{entropy} variable contains either 0 if the return code of \texttt{random\_get} is different of zero, or a pointer to the generated randomness otherwise.
We can see in the code that if \texttt{random\_get} returns an error code, the reference value is set to a deterministic value.
Indeed, dereferencing the \texttt{\_\_stack\_chk\_guard} variable will always return the same value, as in WebAssembly there is no randomization of the memory addresses.
Each variable is thus stored at the exact same memory location at each execution.
This location can be extracted directly from the WebAssembly binary before execution.
If the attacker does not have access to the WebAssembly binary, it can also be easily bruteforced.
Along with the code, this value can be multiplied with the $1103515245$ constant to obtain the reference value.
This means SSP in WebAssembly is fragile, as a failure to get randomness through the \texttt{random\_get} function will systematically result in a predictable reference value.

However, we do not know in which situations the \texttt{random\_get} function may return an error code.
This does not depend on the \texttt{wasi-libc} source code, and as such we need to consider the software used to provide randomness to WebAssembly as an indirect part of the Stack Smashing Protection.
The implementation of how the \texttt{random\_get} WASI function is providing randomness depends on the WebAssembly runtime, and subsequently the host.
As such, the Stack Smashing Protection in WebAssembly is inherently dependent on the runtime implementation.

In order to further assess the robustness of the SSP implementation in WebAssembly, we need to evaluate the implementation of runtimes.
Evaluating thoroughly runtimes and hosts is impractical due to the great amount of possibilities.
In order to get a glimpse of the attacking possibilities, we choose to evaluate the most common WebAssembly standalone runtimes on a classic Linux machine.
We evaluate the robustness of the implementations using two methods:

\begin{enumerate}
  \item[\textbf{M1}] We block the \texttt{getrandom} Linux syscall that is commonly used to acquire randomness on Linux.
  \item[\textbf{M2}] In addition to \textbf{M1}, we block all access to the \texttt{/dev} folder on Linux, which contains the other common source of randomness, the \texttt{/dev/urandom} block device.
\end{enumerate}

To assess whether the implementations of various Linux runtimes are correctly providing randomness, we use a simple C program compiled to WebAssembly, that displays the value of the reference value.
This value is supposed to change at each execution of the program.
If the value repeats itself throughout several executions, it means that the implementation is not able to provide randomness and is returning an error with \texttt{random\_get}.

We describe here the methodology used to assess the robustness of runtimes regarding their implementation of \texttt{random\_get}.
The experimentation was made on the latest version available of the most popular standalone WebAssembly runtimes at the time of the experiment.
The machine used was running Arch Linux with a Linux kernel of version 6.8.5, but the experiment is not dependent on the operating system nor the Linux version up to a point, and should be reproducible in any recent Linux distribution.

The steps are detailed below. They suppose that the runtimes to evaluate are already installed, along with the \texttt{wasi-sdk} in \texttt{/opt}. 
The mentioned files (\texttt{poc.c} and \texttt{seccomp.c}) are made available in our GitHub repository.\footnote{https://github.com/mh4ck-Thales/Robust-SSP-in-Wasm}

\begin{enumerate}
    \item Compile \texttt{poc.c} with \texttt{/opt/wasi-sdk/bin/clang -fstack-protector-all poc.c -o poc.wasm}
    \item Compile \texttt{seccomp.c} with \texttt{clang seccomp.c -lseccomp -o seccomp} 
    \item For assessing \textbf{M1}, run \texttt{./seccomp <tested runtime> poc.wasm}.
    \item For assessing \textbf{M2}:
    \begin{enumerate}
        \item Enter a user namespace with \texttt{unshare -mUr}. You should see that you are now \texttt{root} in the new namespace.
        \item Run \texttt{mkdir /tmp/empty}
        \item Run \texttt{mount --bind /tmp/empty /dev}
        \item Run \texttt{./seccomp <tested runtime> poc.wasm}
    \end{enumerate}
\end{enumerate}

These testing methods are simulating potential attacks, misconfigurations, or other cases.
For example, a WebAssembly runtime in a hardened container may have restricted access to some Linux resources available in the \texttt{/dev} folder.

For each configuration, we execute our test program twice.
If the reference value holds the same value, it means that the implementation is not able to provide randomness, and thus returns an error with \texttt{random\_get}.
This situation is marked with \xmark.
If the reference value holds a different value, it means that \texttt{random\_get} returned randomness.
This does not mean that the provided randomness is secure, merely that the runtime chose to provide randomness and not return an error.
This situation is marked with \cmark.
The results of this evaluation are presented in Table~\ref{security-analysis::random-eval}.

\begin{table}[!ht]
  \centering
  \label{security-analysis::random-eval}
  \begin{tabular}{|c|c|c|c|}
  \hline
  \hspace*{0.1em} Test configuration \hspace*{0.1em} & \hspace*{0.1em} Baseline \hspace*{0.1em} & \hspace*{0.1em} \textbf{M1} \hspace*{0.1em}    & \hspace*{0.1em} \textbf{M2} \hspace*{0.1em} \\
  \hline
  wasmtime (v19.0.1) & \cmark   & \cmark & crash  \\
  wasmedge (v0.13.5) & \cmark   & \cmark & \cmark \\
  wasmer (v4.2.8)    & \cmark   & \cmark & crash  \\
  iwasm (v1.3.2)     & \cmark   & \xmark & \xmark \\
  wasm3 (v0.5.0)     & \cmark   & \xmark & \xmark \\
  wasmi (v0.31.2)    & \cmark   & \cmark & crash  \\
  \hline
  \end{tabular}
  \vspace{1em}
  \caption{Summary of the different configurations w.r.t. the access to random sources}
  \vspace{-2em}
\end{table}

Two runtimes out of the six tested are failing to provide randomness with the situation \textbf{M1}.
In situation \textbf{M2}, the same runtimes are failing to provide randomness, along with three more runtimes that are crashing when trying to provide randomness in this situation.
The remaining runtime is seemingly able to provide randomness.
However, further inspection of the source code is required to ensure the quality of the returned randomness.

Reconsidering the global problem again, we find that the shifting of randomness acquisition from the host (through the libc) to the runtimes may be a problem for the robustness of the SSP implementation.
Most tested runtimes are either unable to provide randomness, triggering \texttt{wasi-libc} to use a predictable value, or are crashing when trying to provide randomness.
One may argue that crashing, at least, does impeach the potential exploitation of weak SSP.
However, a runtime crash is not desirable, especially as \texttt{random\_get} has the possibility to return an error, letting the \texttt{wasi-libc}, and as such the program, handle such a case.

\textbf{P1} is depending both on \texttt{wasi-libc} and on the runtime.
We conclude that the \texttt{wasi-libc} implementation is weak if runtimes are failing to provide randomness, and that several runtimes do in fact fail to provide randomness in some situations.
\textbf{P1} is thus not verified in several of the tested runtimes.

\subsection{Evaluating the SSP reference value location}

In this part, we assess the property \textbf{P2} which states that the location of the reference value must not allow for a bypass of the SSP.
Indeed, if the reference value can be overwritten by a buffer overflow, this can be used to bypass the canary protection.
The attacker just needs to overwrite both the canary and the reference value to the same value.
Two properties can be used to protect against such an attack:

\begin{enumerate}
  \item[\textbf{P2a}] The reference value is not located in a position that is accessible with the overflow of the target buffer.\label{P2a}
  \item[\textbf{P2b}] The memory in which the reference value is located is not writable, or some memory between the buffer and the reference value is not writable.\label{P2b}
\end{enumerate}

In order to assess \textbf{P2a}, we modified the \emph{CookieCrumbler} tool from Bierbaumer et al.~\cite{ifip::18::cookiecrumbler} for WebAssembly with the following modifications:

\begin{itemize}
  \item  The functionalities allowing to check if the range between the buffer and the reference value is writable was removed.
  This functionality could have been used to assess \textbf{P2b}, however its implementation is using signals, which are not supported by WebAssembly. Moreover, this part is useless in WebAssembly, as explained below.
  \item The way \emph{CookieCrumbler} is accessing the memory address of the reference value was adapted to WebAssembly.
  \item The code testing the threads was removed.
  Threads support in WebAssembly is in the process of being standardized, and some WebAssembly runtimes support threads in beta, but the adoption is not wide enough to be studied in this work.
\end{itemize}

\begin{figure}[ht!]
	\centering
	\includegraphics[width=.9\textwidth]{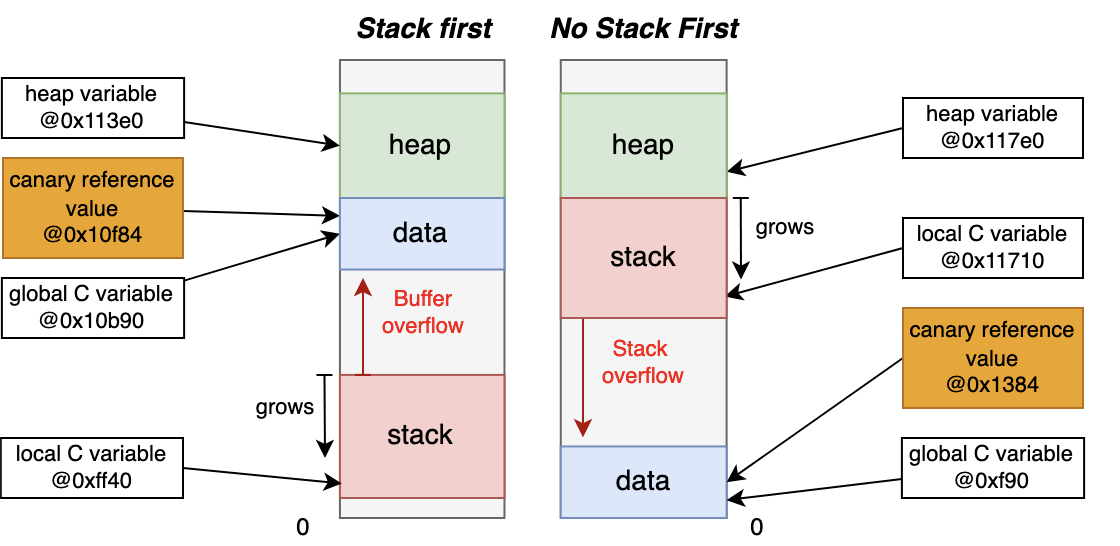}
	\caption{The results of \emph{CookieCrumbler} in the \emph{stack-first} and \emph{no-stack-first} layouts}
	\label{fig::cookiecrumbler}
\end{figure}

We execute \emph{CookieCrumbler} C program compiled with the \texttt{clang} LLVM compiler in both the \emph{stack-first} and \emph{no-stack-first} layouts.
The results are presented in Figure~\ref{fig::cookiecrumbler}.

We draw the following conclusions:

\begin{itemize}
  \item
  A buffer overflow in the \emph{no-stack-first} situation cannot access the reference value, but it is important to note that a stack overflow could.
  \textbf{P2a} is thus verified in the \emph{no-stack-first} layout.
  \item
  With the \emph{stack-first} layout, a buffer overflow from any memory zone, as soon as the overflow is long enough, can overwrite the reference value and bypass the canary protection.
  \textbf{P2a} is thus not verified in the \emph{stack-first} layout.
\end{itemize}

Regarding \textbf{P2b}, the very design of the WebAssembly linear memory makes it impossible to verify this condition.
Indeed, with the lack of memory permissions in WebAssembly, all addresses in the linear memory are writable.
This makes the mapping of the memory containing the reference value as read-only impossible.
Likewise, all the addresses located between the buffer and the reference value are guaranteed to be writable.

\textbf{P2b} is thus not verified in both the \emph{stack-first} and \emph{no-stack-first} layouts.
Consequently, \textbf{P2} is not verified in both layouts.
However, the two layouts are not equal in terms of robustness. 
While the \texttt{stack-first} layout does not verify \textbf{P2} at all, the \texttt{no-stack-first} layout does not allow an overwrite of the reference value with a stack-based buffer overflow.
This layout may still be exploited using another attack primitive alongside the stack-based buffer overflow, but this is a more complex attack.

\subsection{Evaluating quick termination on canary corruption}

This part is assessing if the Stack Smashing Protection mechanism is aborting quickly in case of a canary corruption, i.e.~\textbf{P3}.
If the canary value is corrupted, data in the linear memory is probably corrupted as well.
This means that the program must abort as soon as possible in order to prevent the use of corrupted data.
In all SSP implementations, the detection of canary corruption is made at the end of each function.
Thus, the detection of a memory corruption is bounded by the duration of the execution of the current function.

\begin{figure}
\centering
\includegraphics[width=.9\textwidth]{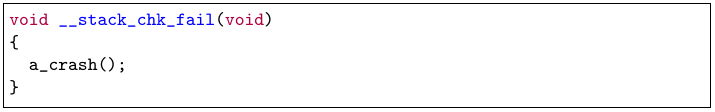}
\caption{The \texttt{\_\_stack\_chk\_fail} C function}
\label{fig:c-stack-chk-fail}
\end{figure}

The abort procedure is implemented in \texttt{wasi-libc}, more precisely in the \texttt{\_\_stack\_chk\_fail} function.
Its source code is shown in Figure~\ref{fig:c-stack-chk-fail}.

\begin{figure}
\centering
\includegraphics[width=.9\textwidth]{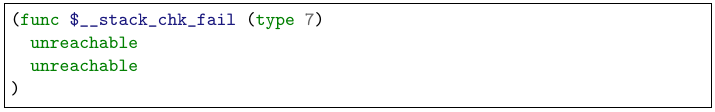}
\caption{Disassembly of the \texttt{\_\_stack\_chk\_fail} WebAssembly function}
\label{fig:wat-stack-chk-fail}
\end{figure}

To verify that the \texttt{a\_crash} function is indeed aborting as soon as possible, we disassemble the compiled \texttt{\_\_stack\_chk\_fail} WebAssembly function to get its assembly code in the WebAssembly Text (WAT) format, shown in Figure~\ref{fig:wat-stack-chk-fail}.

This function is called directly as soon as the corruption is detected.
By inspecting the code, we can see that the function seems to abort the program directly, by executing a WebAssembly \texttt{unreachable} instruction.
Thus, this SSP implementation aborts immediately once the canary value is detected as corrupted.
We conclude that \textbf{P3} is verified.

\subsection{Main findings}

Among the three criteria given to assess the robustness of an SSP implementation, only the quick termination criteria \textbf{P3} is verified by the SSP implementation in standalone WebAssembly.
The criteria on the unpredictability of the canary value \textbf{P1} can be violated in some WebAssembly runtimes which do not crash when access to the host random number generator is impossible.
This lack of randomness can be used to guess the canary value.
The criteria on safe location of canary reference value \textbf{P2} is violated since the WebAssembly SSP reference value is stored in linear memory without protection against as a vulnerable stack buffer.

These weaknesses are exploitable in practice, as our second proof of concept, corresponding to the files ending with \texttt{-ssp} in the artifact repository\footnote{\url{https://github.com/mh4ck-Thales/Robust-SSP-in-Wasm}}, illustrates.

\subsection{Remediation proposals}
\label{proposal}

To fix the weaknesses uncovered in our analysis, we design and implement modifications to the WebAssembly SSP implementation.

\subsubsection{Implementing overwrite protection.}

The WebAssembly linear memory does not allow to store the canary reference value safely, as it may always be overwritten no matter where it is stored.
As a result, it is necessary to store the canary reference value in another WebAssembly memory region.
Moreover, we need to be able to access to this value from the whole WebAssembly module.

Global variables are the only memory mechanism that meet these requirements.
They can only be accessed using WebAssembly instructions, and they are stored in a safe, VM-managed memory.
Thanks to the WebAssembly protections, an attacker cannot execute arbitrary code to try and access the canary reference value.

\subsubsection{Implementing protection against weak randomness.}

Weak randomness can manifest itself in several ways in an SSP implementation.
It can come from the host machine, the runtime, or the library handling SSP.
As a consequence, the SSP implementation should be able to handle all these possibilities.

Sadly, there is no reliable way to prevent against a weak randomness if it is coming from the host or the runtime.
However, if the runtime is correctly implemented, it should return an error with \texttt{random\_get} if it detects that the host or itself is not able to provide strong enough randomness.
The library is then in charge of dealing with the error.

To deal with an error from the \texttt{random\_get} function, the
library may try to call the function later.
However, this is not generally a relevant approach since it often comes from a permanent failure situation.

Developers might be tempted to generate a random value themselves from the library, but they would need to find another source of randomness using WASI, which seems improbable. Falling back on using the current time, despite being a popular idea, is \emph{not} a robust solution.

This is why we believe the only acceptable course of action when \texttt{random\_get} fails is to abort the program during its preamble, thus avoiding running a program with a weak SSP.
While this stance may be controversial on availability and practical considerations, it is the only safe way to enforce security against a weak randomness coming from the host or the runtime.

\vspace*{1em}

We implemented these proposals in the LLVM and \texttt{wasi-libc} projects. This modified toolchain is the one evaluated in the following section.

\section{Evaluation}
\label{evaluation}

In this section, we propose to evaluate the efficiency of our implementation of SSP in WebAssembly.
We use an approach similar to Stiévenart et al.~\cite{qrs::21::stievenart} which compares the execution of programs of the Juliet test suite v1.3~\cite{computer::12::juliet_test}.
The Juliet test suite is a large collection of vulnerability scenarios written in C and organized by MITRE CWE numbers.
In our experiment, we only analyze CWE121~\cite{cwe-121} and CWE122~\cite{cwe-122} tests which respectively correspond to stack-based and heap-based buffer overflows.
We observe the root cause of crashes in the test and classify them in four categories: \emph{silent execution}, \emph{memory fault}, \emph{SSP fault}, \emph{timeout}.
A \emph{silent execution} is an execution which terminates without a crash.
Since all executions lead to an out-of-bound write operation, a silent execution corresponds to a failure to detect a buffer overflow.
A \emph{timeout} occurs as some programs never terminate, which forces us to use a timeout value of $20$ seconds.
A \emph{memory fault} is an execution aborted by a memory fault such as SEGFAULT or SIGBUS.
An \emph{SSP fault} is a crash triggered by the SSP mechanism.

In our experiment, we consider five configurations selected according to two parameters.
The first parameter is whether the binary is a native x86 binary or a WebAssembly binary.
The second parameter is the presence or absence of SSP.
In all configurations, we use LLVM with \texttt{clang} and \texttt{clang++} compilers in version 17.
WebAssembly configurations use the \texttt{wasmtime} runtime and \texttt{wasi-sdk} in version 21.
We focus exclusively on the \emph{stack-first} memory layout after observing that using the default memory layout of LLVM or stack-first yields similar results.

\subsubsection{Observations.}

\begin{figure}[ht]
	\centering
    \vspace*{-.5cm}
	\includegraphics[width=.95\textwidth]{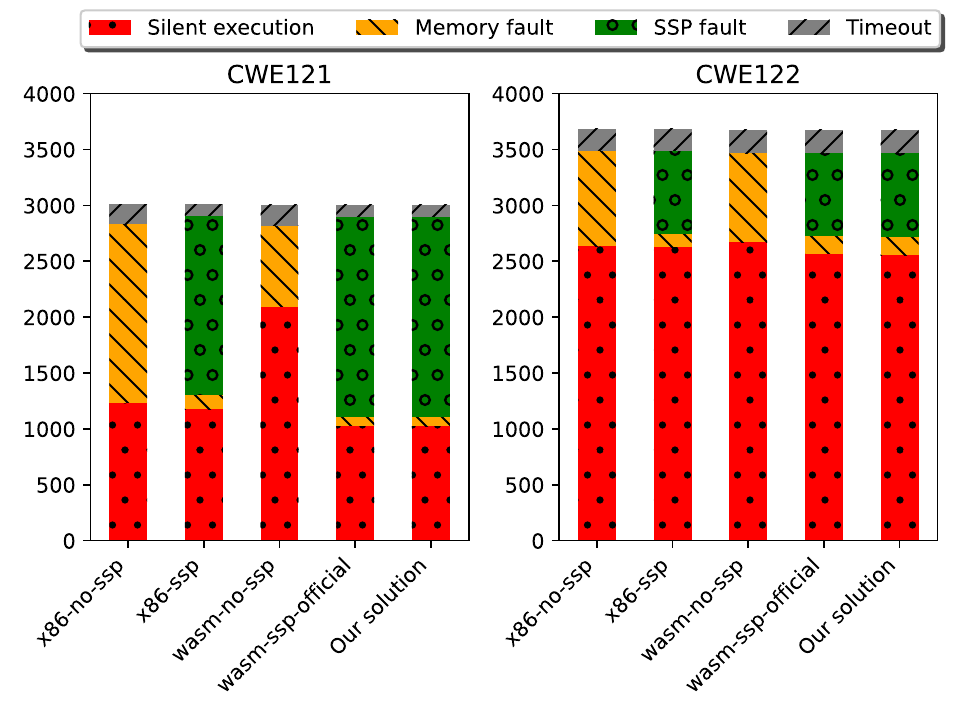}
    \vspace*{-.5cm}
	\caption{Execution outcome of each binary in the Juliet test suite}
	\label{juliet_test}
\end{figure}

The results of our experiment are presented in Figure~\ref{juliet_test}.
For CWE 121, we observe that 24\% of crashes are caused by memory faults for WebAssembly with SSP disabled.
In x86 binaries using SSP, we observe that 53\% of crashes are caused by an SSP fault.
Both the existing implementation and our proposal are able to detect 60\% of buffer overflows.
This proves that our solution is as performant as the original one.

For CWE 122, we observe 22\% of memory faults for WebAssembly with SSP disabled.
x86 with SSP results in 21\% of SSP faults.
Both the existing implementation of SSP in WebAssembly and our proposal are able to detect 20\% of buffer overflows.

The results presented here are consistent with figures reported by Stiévenart et al.~\cite{qrs::21::stievenart}.

\subsubsection{Interpretation.}

First, native and WebAssembly configurations using SSP mitigate more than half stack-based buffer overflows (CWE 121).
This confirms that SSP in WebAssembly is efficient at mitigating stack-based buffer overflows, compared to the situation without protection.
Surprisingly, we observe that some heap-based buffer overflow (CWE 122) of the Juliet test suite crash because of an SSP fault.
This behavior is not expected since a heap overflow grows farther from stack memory, i.e.~from the canary.
We found that all CWE 122 SSP faults occur because the corresponding Juliet tests have been mistakenly tagged as CWE 122, while they are effectively stack-based buffer overflow (CWE 121).
This confirms the expected result that SSP cannot detect heap-based buffer overflows.

Second, our implementation of SSP has the same coverage as the existing implementation.
However, as pointed out in Section \ref{security-analysis-of-webassembly-ssp}, the existing SSP implementation can easily be bypassed.

Third, our implementation is not able to cover the entirety of buffer overflows, in particular a buffer overflow is not detected when the overflow does not reach the canary.
This can happen with small overflows, when e.g.~other variables are allocated between the vulnerable buffer and the top of the stack frame.
However, this defect is common to all SSP implementations.

These results validate the effectiveness of SSP in WebAssembly, and prove that our proposed implementation is as safe and efficient as the existing one.

\section{Conclusion}\label{conclusion}

In this paper, we focused on the mitigation of stack-based buffer overflows in WebAssembly with the Stack Smashing Protection mechanism.
SSP is particularly interesting as it is one of the few binary protections that does not require to modify the WebAssembly specification.

We evaluated the existing implementation of SSP in WebAssembly.
Two weaknesses were identified: the possibility to overwrite the canary reference value and a fragile fallback in case of a random generator failure.

An SSP solution for WebAssembly that mitigates these weaknesses was specified and implemented.
The solution improves the robustness of the existing SSP implementation by proposing secure storage of the canary reference value and a hardened fallback in case of a random generator failure, without any loss of efficiency in detection.

We evaluated our solution and demonstrated that it mitigates a significant portion of stack-based buffer overflows, while being more robust than the already existing one.
This proves the positive impact of this protection on WebAssembly security, leading us to believe that SSP should become a default in all WebAssembly binaries in the future.

The theoretical analysis detailed in this paper is generalizable to all WebAssembly toolchain implementations.
We publish as open-source software the tools used for our analysis, as well as our implementation of SSP.
We hope our work and the related code will be useful to help the community to build safe and secure WebAssembly applications and tooling.

\noindent \textbf{Acknowledgements ---}   
This work has received funding from by the Smart Networks and Services Joint Undertaking (SNS JU) under the European Union's Horizon Europe research and innovation programme under Grant Agreement No 101139067. Views and opinions expressed are however those of the author(s) only and do not necessarily reflect those of the European Union. Neither the European Union nor the granting authority can be held responsible for them.

\bibliographystyle{splncs04}
\bibliography{main}

\end{document}